\documentclass[twocolumn,table,xcolor]{aastex63}
\usepackage[ruled,vlined,norelsize]{algorithm2e}

\sloppypar
\shorttitle{excalibur: a wavelength-calibration model}
\shortauthors{zhao, hogg, bedell, fischer}

\newcommand{\project}[1]{\textsl{#1}}
\newcommand{\name}{\project{excalibur}}
\newcommand{\Name}{\project{Excalibur}}
\newcommand{\acronym}[1]{{\small{#1}}}
\newcommand{\expres}{\project{\acronym{EXPRES}}}
\newcommand{\espresso}{\project{\acronym{ESPRESSO}}}

\newcommand{\eprv}{\acronym{EPRV}}
\newcommand{\lfc}{\acronym{LFC}}
\newcommand{\code}[1]{\texttt{#1}}
\newcommand{\target}{HD 34411}

\newcommand{\mps}{\mathrm{m\,s^{-1}}}

\begin{document}
\title{\Name:
  A Non-Parametric, Hierarchical \\
  Wavelength-Calibration Method for a Precision Spectrograph}

\correspondingauthor{Lily Zhao}
\email{lily.zhao@yale.edu}

\author[0000-0002-3852-3590]{Lily Zhao}
\affil{Yale University, 52 Hillhouse, New Haven, CT 06511, USA}
\affil{Flatiron Institute, Simons Foundation, 162 Fifth Avenue, New York, NY 10010, USA}

\author[0000-0003-2866-9403]{David W. Hogg}
\affil{Center for Cosmology and Particle Physics, Department of Physics, New York University, 726 Broadway, New York, NY 10003, USA}
\affil{Flatiron Institute, Simons Foundation, 162 Fifth Avenue, New York, NY 10010, USA}
\affil{Center for Data Science, New York University, 60 Fifth Avenue, New York, NY 10011, USA}
\affil{Max-Planck-Institut f\"ur Astronomie, Königstuhl 17, D-69117 Heidelberg, Germany}

\author[0000-0001-9907-7742]{Megan Bedell}
\affil{Flatiron Institute, Simons Foundation, 162 Fifth Avenue, New York, NY 10010, USA}

\author[0000-0003-2221-0861]{Debra A. Fischer}
\affil{Yale University, 52 Hillhouse, New Haven, CT 06511, USA}

\begin{abstract}\noindent%
\Name\ is a non-parametric, hierarchical framework for precision wavelength-calibration of spectrographs.  It is designed with the needs of extreme-precision radial velocity (\eprv) in mind, which require that instruments be calibrated or stabilized to better than $10^{-4}$ pixels.  Instruments vary along only a few dominant degrees of freedom, especially \eprv\ instruments that feature highly stabilized optical systems and detectors.  \Name\ takes advantage of this property by using all calibration data to construct a low-dimensional representation of all accessible calibration states for an instrument.  \Name\ also takes advantage of laser frequency combs or etalons, which generate a dense set of stable calibration points.  This density permits the use of a non-parametric wavelength solution that can adapt to any instrument or detector oddities better than parametric models, such as a polynomial.  We demonstrate the success of this method with data from the \textsl{EXtreme PREcision Spectrograph} (\expres), which uses a laser frequency comb.  When wavelengths are assigned to laser comb lines using \name, the RMS of the residuals is about five times lower than wavelengths assigned using polynomial fits to individual exposures.  Radial-velocity measurements of HD~34411 showed a reduction in RMS scatter over a 10-month time baseline from $1.17$ to $1.05\, \mps$.
\end{abstract}

\section{Introduction} 
Precise, radial-velocity programs have been fruitful in finding and characterizing extra-solar planets \citep[e.g.][]{mayor2011, bonfils2013, plavchan2015, butler2017}.  These programs typically make use of spectrographs with resolutions on the order of $50,000-100,000$, which correspond to line widths on the order of $3000\,\mps$.  The state of the art RV precision reached $1\, \mps$ by 2016 \citep{fischer2016}.  The newest generation of instruments aim  to reach $0.1\,\mps$ precision, the required precision to detect terrestrial worlds.  This requires new spectrographs to be calibrated or stabilized to better than $10^{-4}$ of a pixel (assuming that the spectrographs are well sampled).  Two next-generation spectrographs, \expres\ and \espresso, have been commissioned for more than a year and are demonstrating $<0.1 \mps$ instrumental errors and $\sim 0.2\, \mps$ errors on stars \citep{pepe2013, jurgenson2016, blackman2020, petersburg2020, brewer2020, mascareno2020}.

Traditionally, wavelength solutions are constructed by fitting a polynomial to lines from a calibration source in order to describe the relationship between wavelength and pixel for each echelle order \citep{butler1996, lovis2007, cersullo2019}.  In this framework, each calibration image is treated independently.  The returned wavelength solutions worked well at the level of 1 $\mps$ precision.

The move towards 0.1 $\mps$ RV precision, necessitates higher-fidelity calibration data and wavelength models.  These models need to account for high-order spatial variations that can arise from small imperfections in the optics of an instrument and non-uniformity in detector pixel sizes/spacing.  There has been significant effort in using an entire set of calibration images to identify incongruous ThAr lines \citep{coffinet2019} or obtain high-resolution Fourier transform spectra of reference cells \citep{wang2020}.  It has also been found that using multiple polynomials in the dispersion direction, tuned to capture detector defects, better describes the wavelength solution than a single, continuous polynomial \citep{milakovic2020}.

Here, we propose to simplify and improve calibration programs for \eprv\ hardware systems with two practical yet innovative ideas.  The first flows from the fact that calibration sources---which include arc lamps (in some wavelength ranges), etalons, and laser-frequency combs (\lfc s)---illuminate the spectrograph with very stable, very dense sets of lines--almost every location in the spectrograph image plane is surrounded by nearby, useful calibration lines.  This enables use of a calibration methodology that is \emph{non-parametric}, or not defined by a prescribed, analytic function described by a finite number of parameters:  If every point in the spectrograph detector is sufficiently surrounded by nearby calibration lines, the wavelength solution can, for example, be made simply as an interpolation of the calibration data.  The density of lines removes the need to enforce any functional form for the wavelength solution (such as a continuous ninth-order polynomial).  In some ways, this is a generalization of recent work that has demonstrated the efficacy of constructing a wavelength solution as multiple, segmented polynomials \citep{milakovic2020}.  A non-parametric approach will improve calibration accuracy by not forcing the choice of a parametric form that may bias the calibration, especially when the chosen function is inappropriate (as, for example, polynomials are at detector edges).

The second simple idea follows from the observation that most physical systems have only a few dominant degrees of freedom, meaning most spectrographs vary along only a small number of axes in ``calibration space'', or the (very high-dimensional) space of all possible wavelength solutions.  This is particularly true of \eprv\ instruments, which are equipped with stringent environmental stabilizing.  The thermomechanical stability of these instruments reduces the variations they experience to something that can be represented by a low-dimensional framework.  That is, spectrographs, especially stabilized ones, should have few environmentally accessible degrees of freedom.  This renders it inadvisable to fit each calibration exposure or calibrate each science exposure independently.  Instead, all the calibration data (or all the data) should be used to determine the calibration space in which the instrument can and does vary.  Subsequent calibration work then need only determine where in the small, accessible part of calibration space the spectrograph was located for each exposure.

In the context of probabilistic models, this structure is \emph{hierarchical}:  The calibration data are used not just to determine the wavelength solution at one moment, but also to determine the possible \emph{calibration space} of wavelength solutions at all moments.  In statistics, this concept is often described as \emph{de-noising}:  we can improve calibration by recognizing that every calibration exposure contains information about every other calibration exposure.  Thus, every exposure can be improved (i.e., de-noised) with information from every other exposure.

The method we propose here---\name---embodies these ideas.
It is a non-parametric, hierarchical, data-driven method to generate a wavelength model.  By being non-parametric, it delivers enormous freedom to the wavelength solution to match or adapt to any instrument or detector oddities.  By being hierarchical, it restricts that freedom tremendously, but it does so appropriately for the empirically determined variations in a spectrograph.

The method \name\ is designed for temperature-controlled, fiber-fed spectrographs with good calibration sources, such as laser-frequency combs, or etalons.  We have in mind \eprv\ instruments and \eprv\ science cases, primarily because the need for good wavelength calibration is so great in this field.  Irregardless, we expect \name\ to have applications for other kinds of spectrographs in other contexts.  \Name\ should be applicable to other spectrograph with low-dimensional variability, though the precision of the returned wavelengths will depend on the available calibration sources (more discussion in Section \ref{sec:others} below) .

\section{Method} \label{sec:method}
\begin{figure*}[t]
\centering
\includegraphics[width=\textwidth]{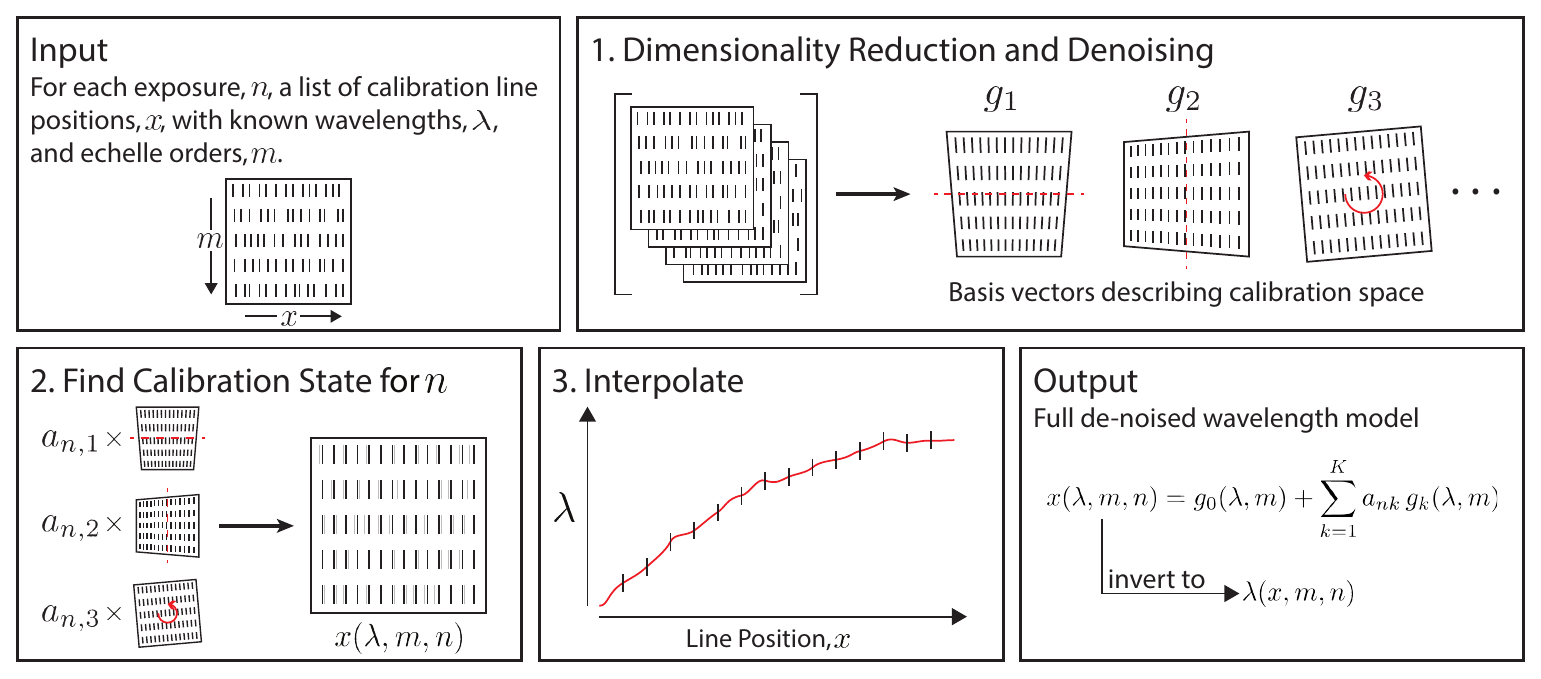}
\caption{A cartoon representation of the \name\ method, as described in Section \ref{sec:method}.  We exaggerate variations in measured line position, changes in calibration space, and interpolation deviations for clarity.  In step one, dimensionality reduction and denoising (\textsection \ref{sec:denoising}), the complete set of line positions for all exposures is analyzed to return a set of $K$ basis vectors, $g_n$, which represent different ways the spectrograph calibration changes.  These basis vectors span the $K$-dimensional calibration space of the spectrograph, which includes all possible wavelength solutions.  In step two (\textsection \ref{sec:interp_time}), the amplitude of each basis vector, $a_{n,k}$, is interpolated to return the calibration state for a specific science exposure, returned as a set of de-noised calibration lines.  The assigned wavelengths of these de-noised line positions are then interpolated onto other pixels in step three (\textsection \ref{sec:interp_wsol}) to construct a full wavelength model that returns wavelength as a function of detector position $x$ and echelle order $m$.}
\label{fig:cartoon}
\end{figure*} 

The \name\ method is designed to take many calibration images, each containing a series of calibration lines with known wavelengths and well-fit detector positions, and de-noise and interpolate this information into a full wavelength model applicable to all exposures taken with the instrument.  It operates on two core ideas; The wavelength solution should be allowed flexibility, but it lives in a very low-dimensional calibration space where the degrees of freedom are set by the limited kinematics of the spectrograph hardware.  \Name\ therefore assumes that the space of possible calibration states for an instrument is low-dimensional but assumes very little about the forms of those states.

\Name\ also assumes dense-enough calibration line coverage with well-fit line centers to provide sufficient constraints on an interpolated wavelength solution across an echelle order.  Upstream errors in line center positions may propagate through \name\ wavelength models.  The required line density is dependent on the required precision of the returned wavelength model; larger spacing between lines offer less constraint and are likely to return worse wavelengths.  We revisit and quantify these conditions in Section \ref{sec:others}.

Wavelength calibration is usually posed in the following way: Given an exposure $n$, and echelle order $m$, there is a relationship between
the two-dimensional $(x,y)$-position on the detector and the
wavelength $\lambda$
\begin{equation}
\lambda(x,y,m,n) = f(x,y,m;\theta_{n})
\quad ,
\label{eq:wsol}
\end{equation}
where $\theta_{n}$ represents the parameters describing the wavelength solution for a given exposure.

Classically, pipelines employ polynomials to construct smooth wavelength solutions for each exposure.  For example, the \expres\ pipeline sets the function $f(x,y,m;\theta_{n})$ from Equation \ref{eq:wsol} to a 2D, 9\textsuperscript{th}-order polynomial, where $\theta_{n}$ represents the polynomial coefficients, $c_{nij}$, unique to each exposure $n$ \citep{petersburg2020}.
\begin{equation}
\lambda(x,m,n) = \sum_{i=0}^9\sum_{j=0}^9 c_{nij}\, x^i\,m^j + \mathrm{noise}
\quad ,
\label{eq:poly_wsol}
\end{equation}
Here, the y-dependence is dropped as, in our framing, this dependence is carried by spectral order $m$.  The line position on the detector is therefore uniquely identified by echelle order $m$ and pixel in the dispersion direction, $x$.  The coefficients $c_{nij}$ are interpolated from the time of calibration exposures to time $t_n$ of a science exposure $n$ by a third-order polynomial with respect to time.  This third-order polynomial is evaluated at the time of non-calibration, science exposures to re-construct the coefficients for a 2D, 9th-order polynomial wavelength solution for that exposure.  Each calibration image obtains its $c_{nij}$ coefficients independently.

Given a stabilized instrument with low degrees of freedom, however, the calibration of any image can be reliably informed by the calibration of every other image.  The calibration data themselves can be used to develop a low-dimensional basis for expressing the space of all possible calibrations for a spectrograph with few degrees of freedom.

If the space of all calibration possibilities is in fact $K$-dimensional (where $K$ is a small integer, i.e 2 or 8 or thereabouts), and if the calibration variations are so small that they can be linearized, then the function $f(x,m;\theta_{n})$ from Equation \ref{eq:wsol} should be low-dimensional.  In \name, we transpose the calibration model---making the position $x$ a function of $\lambda$---into the following form
\begin{equation}
x(\lambda,m,n) = g_0(\lambda,m) + \sum_{k=1}^K a_{nk}\,g_k(\lambda,m)
\quad ,
\label{eq:excl_wsol}
\end{equation}
where
$g_0(\lambda,m)$ is the fiducial or mean or standard calibration of the spectrograph,
the $a_{nk}$ are $K$ scalar amplitudes for each exposure $n$, and the $g_k(\lambda,m)$ are basis functions expressing the ``directions'' in calibration space that the spectrograph can depart from the fiducial calibration.  The resultant $x(\lambda,m,n)$, a list of calibration line positions for a given exposure, can be regarded as the calibration state of the spectrograph for that exposure.  When this calibration structure is used to deliver a wavelength solution, $x(\lambda,m,n)$ can be inverted back into $\lambda(x,m,n)$ to recover wavelengths for each detector position $x$ and echelle order $m$ (see Figure \ref{fig:cartoon}).

The challenge is to learn these basis functions, $g_k(\lambda,m)$, from the data and get the $K$ amplitudes, $a_{nk}$, for every exposure $n$.  There are many ways to discern the basis functions.  In this paper, we present one implementation of \name\ using principal component analysis (PCA) \citep{pearson1901, jolliffe2016}.  A PCA is justifiable in the limit where exposures have very high signal-to-nose ratio, as is usually the case with typical calibration images.  There are many alternatives to PCA for this dimensionality reduction; we return to this point in Section \ref{sec:choices} below.

\subsection{Dimensionality Reduction: De-Noising of Calibration Frames} \label{sec:denoising}
\Name\ will use calibration images to determine 1) the space in which an instrument varies and 2) where in the accessible calibration space the spectrograph existed for each exposure.  For each calibration exposure, $n$, \name\ requires a full list of lines, $(\lambda,m)$ that are expected to appear in each calibration exposure.  Each line is uniquely defined by a combination of echelle order, $m$, and ``true'' or theoretical wavelength, $\lambda$.  There are many strategies for identifying calibration line positions and matching them to their assigned wavelengths; this problem is left out of scope for this work.  

\Name\ assumes that line positions have been identified ``correctly,'' as in the position of a calibration line is determined the same way as the position of a stellar line when extracting RVs.  This also inherently assumes that the calibration lines are not subject to an effect that the science exposures are not, for example differences in charge transfer inefficiency, non-linearities in the PSF, etc.  We caution that systemic errors or large uncertainties in fitting line positions easily propagate through to biases in the wavelength models returned by \name.  For more discussion, see Section \ref{sec:discussion}.

For each exposure, $n$, every line, $(\lambda,m)$, has an associated fitted (measured) detector position, $x(\lambda,m,n)$, for example x-pixel in an 2D extracted echelle order.  Fitted line centers that are missing from an exposure (e.g. because the fit failed due to noise, the line is not in its usual echelle order, etc.) can be assigned a \code{NaN} (hardware not-a-number) for that exposure instead.  Let there be $P$ lines per exposure.  \Name\ reads in a $N \times P$ matrix of line positions for each of $P$ lines for each of $N$ exposures.

The mean of measured line position over the set of calibration exposures represents the fiducial, or standard calibration of the spectrograph, $g_0(\lambda,m)$.  In this implementation of \name, principal component analysis is performed on the difference between this fiducial calibration and each individual line position.  The returned principal components serve as basis functions,  $g_k(\lambda,m)$, expressing the possible deviations of the spectrograph from this fiducial calibration.  The magnitude of each principal component for each exposure, $a_{nk}$, represents the scalar amplitude of these deviations for each exposure.  \Name\ then uses a small number, $K$, of principal components to reconstruct a de-noised version of the line positions as formulated in Equation \ref{eq:excl_wsol}.

Missing line center measurements, which were previously marked by \code{NaN}s, are replaced with de-noised estimates.  This is done iteratively until the estimates of missing line centers change by less than 0.01\%.  This process can be repeated on line centers deemed as outliers by some metric, to account for lines that may have been mis-identified or mis-fit.  The principal components from the final iteration are used to define the spectrograph's calibration space, while the associated amplitudes for each component pinpoint where in that calibration space the spectrograph is located for each calibration exposure.

\begin{algorithm}
\SetAlgoLined
\KwData{line positions $x(\lambda,m,n)$ for each exposure $n$, with wavelengths $\lambda$ and echelle orders $m$}
\KwResult{Basis vectors of the low-dimensional calibration space $g_k(\lambda,m)$ and location of exposures in calibration space expressed by amplitudes $a_{n,k}$}
\While{change in missing or outlier line centers $>$ 0.01\%}
{
	$g_0(\lambda,m) = \overline{x(\lambda,m,n)}$\;
	using Singular-Value Decomposition, find $U, \Sigma, V$ s.t. $U\Sigma V^* = (x(\lambda,m,n)-g_0(\lambda,m))$\;
	let $a_{n,k} = U\cdot \Sigma$ and $g_k(\lambda,m) = V$\;
	$x(\lambda,m,n) = g_0(\lambda,m) + \sum_{k=1}^K a_{nk}\,g_k(\lambda,m)$ for $x(\lambda,m,n)$ = \code{NaN} where $K$ is a a small integer
	}
\caption{Dimensionality Reduction and De-Noising}
\end{algorithm}

\subsection{Interpolating Calibration Position}
 \label{sec:interp_time}
In \name, the amplitude, $a_{n,k}$, of each principal component is interpolated to determine the calibration state of the spectrograph.  For example, the amplitude can be interpolated with respect to time to recreate the calibration state of the spectrograph at different times.  The choice of what to interpolate against depends on the dominant contribution to variation in the calibration of the instrument.

In the implementation of \name\ presented here, the amplitudes of the principal components are interpolated linearly with respect to time.  This is discussed more in Section \ref{sec:choice_avt}.  Let a prime denote values related to a science exposure $n'$ for which we want wavelengths.  We use linearly interpolated magnitudes, $a_{n',k}$ at time $t_{n'}$ to construct the calibration state of the spectrograph for that point in time.  Using interpolated amplitudes, $a_{n',k}$, and the basis vectors, $g_k(\lambda,m)$, returned by the de-noising process, a new set of calibration lines, $x'(\lambda,m,n')$ can be constructed for any exposure as formulated in Equation \ref{eq:excl_wsol}.

\subsection{Interpolating a Wavelength Solution} \label{sec:interp_wsol}
From the de-noising step, \name\ can now construct a set of calibration lines, $x'(\lambda,m,n')$ for any exposure $n'$.  To construct a wavelength solution, we invert $x'(\lambda,m,n')$ to $\lambda(x',m,n')$ by interpolating known wavelengths of the calibration lines over detector position.  For instance, interpolating the known wavelengths vs. line centers onto every integer $x$ will generate wavelengths for each pixel in an echelle order.

After experiments, we found that a cubic-spline interpolation that enforces monotonicity, such as a Piecewise Cubic Hermite Interpolating Polynomial (PCHIP) interpolator, works well for interpolating wavelengths onto pixels.  A cubic spline allows for more flexibility than a parameterized function, while the enforced monotonicity allows the wavelength solution, $\lambda(x',m,n')$, to be invertible and prevents spurious deviations that may befall a cubic spline.  Choices in interpolation scheme, $K$, and other tests are further discussed in Section \ref{sec:choice_wvp}.

\begin{algorithm}
\SetAlgoLined
\KwData{the fiducial calibration of the spectrograph $g_0(\lambda,m)$; magnitudes of the principal components for each exposure $a_{n,k}$; basis vectors spanning the calibration space of the spectrograph $g_k(\lambda,m)$; }
\KwResult{Wavelengths for detector positions $x'(m,n')$ of exposure $n'$ with time $t_{n'}$, where the prime is used to denote values relevant to this new exposure}

Find $a_{n',k}$ by interpolating $a_{n,k}$ with respect to $t_{n'}$\;
$x'(\lambda,m,n') = g_0(\lambda,m) + \sum_{k=1}^K a_{n'k}\,g_k(\lambda,m)$ where $K=6$\;
\For{each unique $m$}{
	interpolate $\lambda$ with respect to $x'(\lambda,m,n')$ onto pixels $x'(m,n')$\;
	}
\caption{Generating Wavelength Solution}
\end{algorithm}

The implementation of \name\ described here is hosted on GitHub\footnote{\url{https://www.github.com/lilyling27/excalibur}}.

\section{Data} \label{sec:data}
We tested \name\ using data from \expres, the EXtreme PRecison Spectrograph.  \expres\ is an environmentally-stabilized, fiber-fed optical spectrograph with a median resolving power $R=\lambda/\delta_{\lambda}=\sim137,000$, over a wavelength range of $390-780\, nm$ \citep{jurgenson2016, blackman2020}.  \expres\ has two different wavelength calibration sources, a thorium argon (ThAr) lamp and a Menlo Systems laser frequency comb (LFC).  LFCs are unique in that the wavelengths of their emission lines are stable and exactly known at pico-meter accuracy \citep{wilken2012, molaro2013, probst2014}.

Rather than using a simultaneous calibration fiber, two to three LFC exposures are obtained roughly every 30 minutes while the telescope is slewing to new targets.  ThAr exposures are taken at the beginning and end of each night.  All calibration data are taken through the science fiber, so that calibration light travels down the same optical pathway and is projected onto the same pixels as the science observations.  Light passes through a pupil slicer and double scrambler before being injected into a rectangular fiber, which is fed through a mechanical agitator to ensure modal mixing\citep{petersburg2018}.

LFC lines cover echelle orders 84-124, which contain approximately 19200 calibration lines.  Though our results are primarily based on work with LFC data, there will be some discussion of applications to arc lamps below.   ThAr lines cover all 86 extracted orders of \expres\ (echelle orders 75-160), which include approximately 5300 lines.  For both the LFC and ThAr data, lines that appeared in less than 60\% of exposures were not included in the analysis.  Similarly, exposures with more than 60\% of expected lines missing were cut from the analysis.  A list of echelle orders $m$, line wavelengths $\lambda$, and pixel positions $x$ were calculated by the EXPRES pipeline \citep{petersburg2020} for every line of every exposure and read into \name.

Line positions from the \expres\ pipeline are generated as follows.  A ThAr wavelength solution is generated from each ThAr exposure using the IDL code \texttt{thid.pro}, developed by Jeff Valenti.  This code identifies ThAr lines by matching lines in an exposure against a line atlas.  Line matching is carried out in an automated and unsupervised way with a Levenburg-Marquardt minimization routine.  Once each line's position is identified and matched to a wavelength from the line atlas, a sixth-order, 2D polynomial is fit over pixel location $x$, echelle order $m$, and scaled wavelength $m\lambda$ (wavelengths are scaled in order to distinguish lines that may appear in more than one order).
 
 Flat-relative, optimally extracted LFC data is background-corrected using a univariate spline.  Each peak in an echelle order is then fit with a Gaussian.  The mean of this fitted Gaussian to a single peak is taken to be the center of the line.  For each line, the ThAr wavelength solution is used to estimate the mode number of a line.  The precise wavelength is then calculated using
 \begin{equation}
 f_n = n \times  f_r + f_0
 \label{eq:lfc}
 \end{equation}
 where the repetition rate, $f_r$, is known from the design of the LFC, and the offset frequency, $f_0$, has been determined by Menlo Systems, the manufacturer of the LFC.
 
In order to comfortably satisfy the assumption that there exists only low-order variation, which is needed for \name, we used exposures from after the LFC stabilized following servicing in summer 2019, where the photonic crystal fiber was replaced and the polarization was changed to shift the wavelength range of the LFC redwards.  In the results presented here, we use 1227 LFC exposures and 78 ThAr exposures taken between October 14 and December 18, 2019 on 29 unique nights.

\section{Tests}\label{sec:tests}
We perform a series of tests to validate the performance of \name\ and benchmark \name -generated wavelengths against wavelengths generated by a classic, non-hierarchical, parametric method.  To implement training-validation tests, we leave out a subset of calibration lines with known wavelengths as a ``validation'' sample, generate wavelengths for these lines using the remaining data, and compare the predicted wavelength to the assigned wavelength of each line.  This inherently folds in errors in the measured line center of each calibration line, but this contribution to the residuals will be the same across all tests.

To assess the classic, polynomial-driven method of wavelength calibration, we take each LFC exposure and separate the lines into even- and odd-indexed lines.  We then construct a wavelength solution using only the odd-indexed lines and use  that wavelength solution to predict wavelengths for the even-indexed lines; i.e. a polynomial is fit to just the odd-indexed lines and then evaluated at the detector positions of the even-indexed lines (see Equation \ref{eq:poly_wsol}).  We then generate a wavelength solution using only the even-indexed lines and use it to predict wavelengths for the odd-indexed lines.

To test the interpolation step of \name\ (\textsection \ref{sec:interp_wsol}), we employed \name\ on all LFC exposures with odd-indexed lines removed.  The resultant basis vectors, $g_k(x,y,m)$,  and amplitudes, $a_{nk}$, are therefore only informed by the even-indexed lines of each LFC exposure.  We then predict wavelengths for the odd-indexed lines that had been excluded and compare these predictions to their assigned wavelengths.  This allows us to test how accurately an interpolated wavelength solution can predict wavelengths.

To test the denoising step of \name\ (\textsection \ref{sec:denoising}, \textsection\ref{sec:interp_time}), we employed \name\ on a randomly selected 90\% of all LFC exposures.  This means the basis vectors, $g_k(x,y,m)$,  and weights, $a_{nk}$, were constructed using only information from 90\% of all exposures.  We used the results to predict wavelengths for all the lines in the remaining 10\% of calibration exposures.  This allows us to test how well we can pinpoint the calibration state of the spectrograph using \name.

The polynomial and interpolation tests remove the same 50\% of lines from each exposure while the denoising test completely removes a randomly selected 10\% of calibration exposures and their associated line position measurements.  Errors from interpolation will be localized, extending only to neighboring lines.  We therefore aggressively remove every other line to ensure we are capturing these local effects.  The PCA denoising, on the other hand, folds in information of all lines from all exposures.  Here, it is sufficient to completely remove 10\% of exposures, a traditional training/validation fraction.  Since the information being removed varies between each test depending on its focus, we present the results per line, treating each line as an independent test.

The residuals of a wavelength solution represent the difference between the wavelength solution evaluated at the line position of a calibration line, and the assigned theoretical wavelength (i.e. from Equation \ref{eq:lfc} for LFC lines) on a line-by-line basis in every exposure.  The reported RMS of a wavelength solution is therefore the per-line RMS, i.e.
\begin{equation}
RMS/line \: [m\,s^{-1}] = \sqrt{\sum_{n=1}^N\sum_{p=1}^P\frac{[ \frac{(\lambda_{n,p,pred.} - \lambda_{p,theory})}{\lambda_{p,theory}} \times c ]^2}{N \times P}}
\label{eq:rms}
\end{equation}
where $\lambda_{p,theory}$ is the theoretical wavelength for line $p$, $\lambda_{n,p,pred.}$ is the wavelength predicted by the constructed wavelength solution for line $p$ in exposure $n$, and residuals from all $P$ lines from all $N$ exposures are used, for a total of $N \times P$ lines.  The difference in wavelength is converted to units of $\mps$, a more intuitive metric for \eprv\ work.

\subsection{Results}
\begin{figure}[b]
\centering
\includegraphics[width=.45\textwidth]{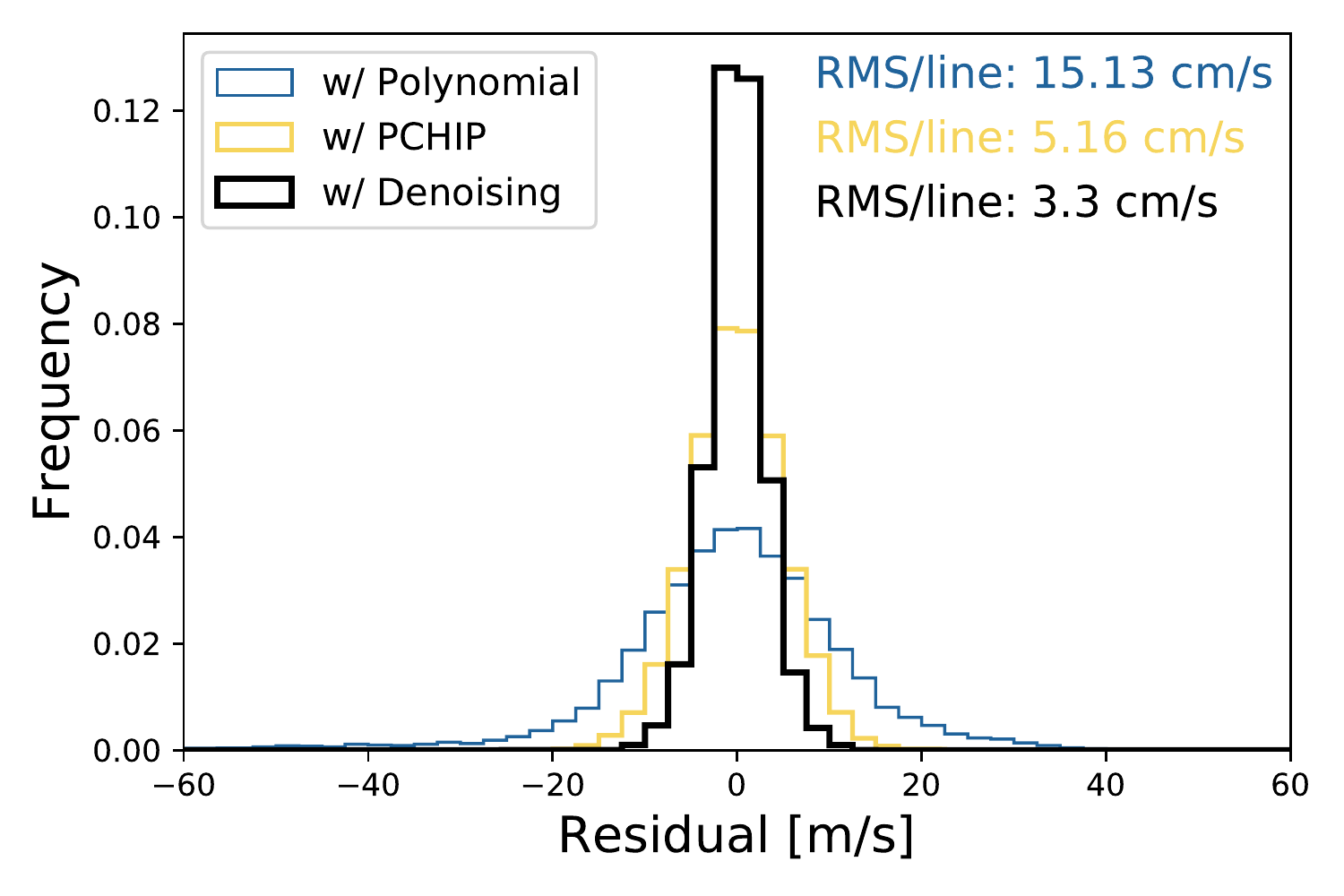}
\caption{Difference in predicted and theoretical wavelengths for the wavelength calibration tests described in Section \ref{sec:tests}.  The per line RMS as defined in Equation \ref{eq:rms} is given in the top-right corner in each method's corresponding color.  Incorporating denoising returns the smallest spread in residuals.}
\label{fig:testHists}
\end{figure}

Histograms of the per-line residuals for each of the above described polynomial, interpolation, and denoising tests (respectively) are shown in Figure \ref{fig:testHists}.  Note that the spread in residuals is much smaller for both the denoising and interpolation tests relative to the results of the polynomial wavelength solution.

\begin{figure*}[t]
\centering
\includegraphics[width=\textwidth]{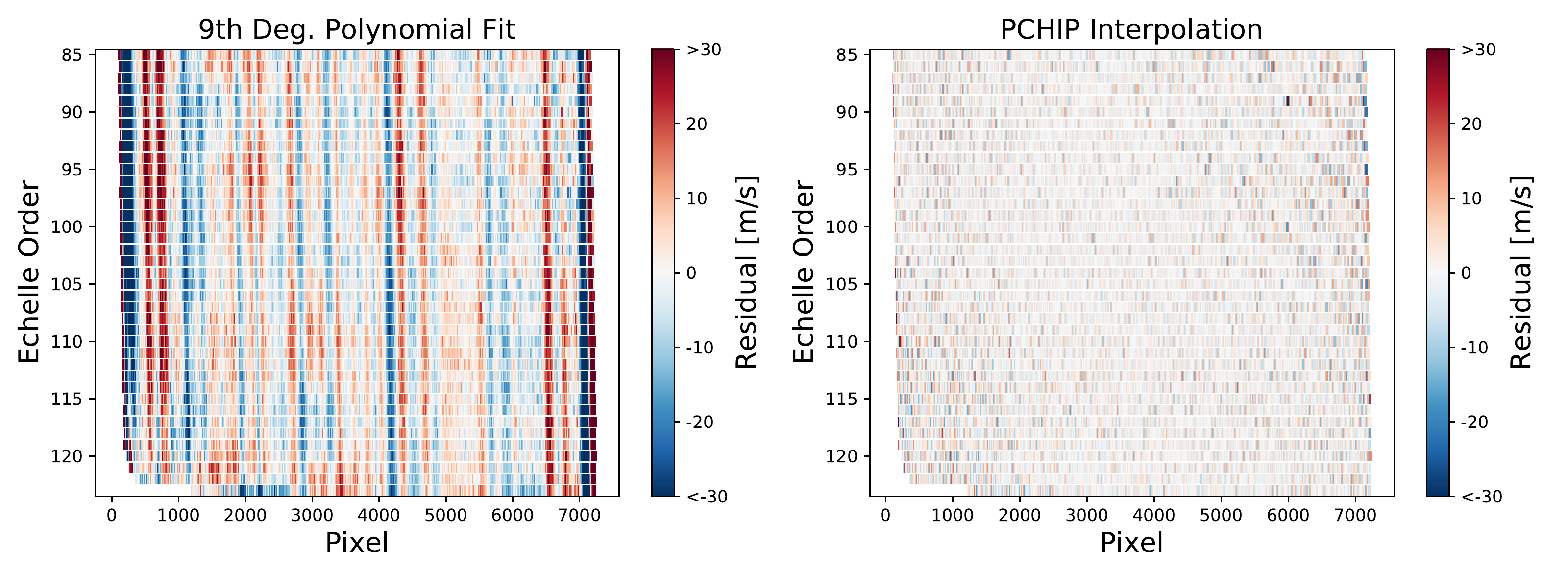}
\caption{Residuals of a single LFC exposure plotted with respect to detector position (as defined by echelle order and x-pixel) for parametric (top) and non-parametric (bottom) wavelength calibration methods.  Each line is colored by the difference between the predicted wavelength and the theoretical wavelength for each line, given in units of $\mps$.  High-order structure, i.e. vertical stripes and patchiness, is apparent in the residuals to a polynomial wavelength solution, which assumes smoothness.}
\label{fig:resid2d}
\end{figure*}

The per-line residuals from the denoising test also exhibit smaller spread than interpolation alone.  This suggests that the spectrograph truly is accurately represented by a low-dimensional model.  Recreating line positions using this model gives better line position estimates than treating each exposure independently.  The low-dimensional model does not incorporate noise from individual line measurements.  Returning more precise, denoised line positions results in smaller per-line residuals.

\Name -generated wavelengths also exhibit less structure in the returned residuals.  For a randomly selected example LFC exposure, Figure \ref{fig:resid2d} plots each line with respect to its echelle order (y-axis) and x-pixel on the detector (x-axis) colored by the difference between the predicted and theoretical wavelength for that line in units of $\mps$.

The residuals of the classic, polynomial wavelength solution is shown in the top plot of Figure \ref{fig:resid2d}.  There is a lot of vertical structure and some hints of a periodic, diagonal structure as well.  The residuals of the interpolation test for the same exposure is shown in the bottom plot of Figure \ref{fig:resid2d}.  There is no coherent structure here and smaller residuals. 

This shows how the flexibility of an interpolated model can account for high-order instrument or detector defects, which emerged as structure in the residuals of the classic, smooth, polynomial-driven wavelength solution.  This same flexibility may similarly allow interpolated wavelength solutions to account for position errors in pixel image blocks for different detectors depending on how the interpolation is framed \citep{fischer2016, milakovic2020}.

\begin{figure*}[t]
\centering
\includegraphics[width=\textwidth]{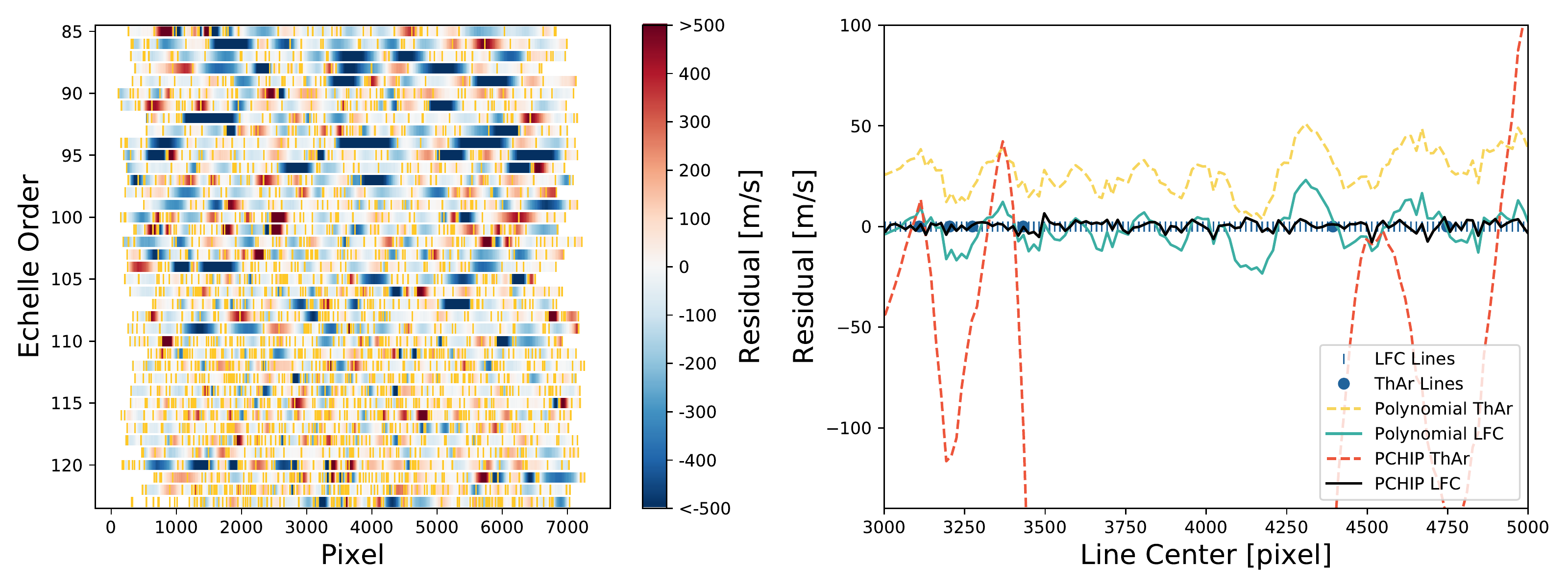}
\caption{Residuals when using ThAr lines to predict wavelengths for LFC lines.  \textbf{Left}: residuals for a single LFC exposure plotted with respect to detector position and colored by residual, as in Figure \ref{fig:resid2d}.  The positions of ThAr lines are over-plotted in yellow.  In general, residuals are greater between ThAr lines with greater separation.
\textbf{Right}: Comparison of polynomial and interpolated wavelength solutions using either just ThAr lines or LFC lines for a subset of echelle order 94.  The shape of the residuals from a polynomial fit are similar whether using ThAr lines or LFC lines.  A PCHIP interpolated wavelength model guided by LFC lines returns the smallest residuals.}
\label{fig:waveResids}
\end{figure*}

Though the interpolated wavelength solution returns lower, less-structured residuals than the polynomial wavelength solution when guided by LFC lines, the flexibility of an interpolated wavelength solution can result in much worse residuals when not properly constrained, for example in regions between widely separated calibration lines.  The left plot of Figure \ref{fig:waveResids} shows the residuals when ThAr calibration lines, which are much fewer and less regularly-spaced than LFC lines, are run through \name\ and used to predict wavelengths for the (completely independent) LFC exposures taken during the same range of time.  Over-plotted in yellow are the positions of the ThAr lines. 

Note that running \name\ informed by only ThAr lines cannot be regarded as a direct comparison to the LFC runs, as the increased uncertainty and variability in ThAr line positions alone makes the resultant wavelength predictions an order of magnitude worse, hence the different scale of the colorbar in the left plot of Figure \ref{fig:waveResids} as compared to Figure \ref{fig:resid2d}.  All the same, the residuals are in general worse where lines are further apart (for example, in redder echelle orders) than where lines are denser.

Figure \ref{fig:waveResids} (right) plots the residuals for a subset of order 94 for both a polynomial-based method and a PCHIP-based method guided by either ThAr lines or LFC lines.   The PCHIP model with ThAr lines (orange, dashed curve) returns huge residuals between two widely-separated ThAr lines that extends out of frame.  The classic, polynomial fit exhibits similar residuals in both amplitude and shape regardless of whether the set of ThAr lines or LFC lines are used.  An interpolated wavelength solution using LFC lines (black, solid curve) exhibits the lowest residuals.

The move to an interpolated wavelength solution is driven by the assumption that a high density of calibration lines allows for more freedom in the resultant wavelength solution.  This freedom allows the wavelength solution to more accurately ascribe wavelengths.  This flexibility, however, is no longer justified in the regime where there are large separations between calibration lines, as this no longer provide sufficient constraint on the interpolated wavelength solution, as is the case in some regions of a classic ThAr lamp.

\subsection{Impact on Radial Velocity Measurements}\label{sec:test-rv}

\begin{figure*}[t]
\centering
\includegraphics[width=\textwidth]{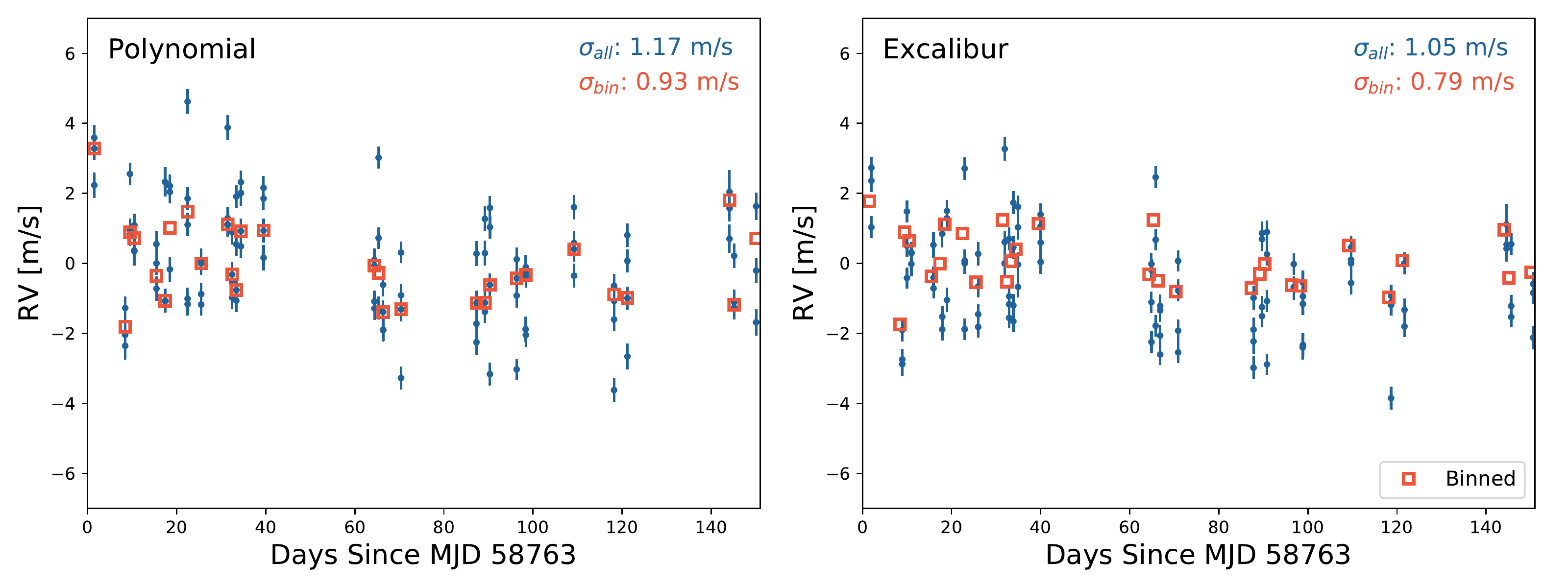}
\caption{HD 34411 RVs measured with \expres.  We zoom in on all but 8 of the exposures, which were taken over 150 days later.  All RVs are shown in blue; nightly binned RVs are over-plotted as orange squares.  The RMS of the data set and the RMS of the binned values are given in the top-left corner.  \textbf{Left}: RVs derived using a polynomial-based wavelength solution.  \textbf{Right}: RVs derived using wavelengths from the implementation of \name\ presented in this paper.}
\label{fig:rvs}
\end{figure*}

We tested \name -generated wavelengths on RV measurements using \expres\ observations of \target, which are presented in Table \ref{tab:rv}.  \target\ is a G0V star.  It is $4.8\, Gyr$ old and relatively quiet ($log R'_{HK} = -5.09$) \citep{brewer2016}.  Because \target\ has no known planets and should have a smaller contribution from stellar signals, it is a good star to test the effects of the wavelength calibration on the RMS of the returned RVs.  We use 114 observations of \target\ taken between October 8, 2019 and March 5, 2020 with SNR 250.  Radial-velocity measurements were derived using a chunk-by-chunk, forward-modeling algorithm ran by the \expres\ team \citep{petersburg2020}.

Figure \ref{fig:rvs} compares the resultant RVs when using a classic, ninth-degree polynomial wavelength solution and an \name -generated wavelength model.  Using \name -generated wavelengths reduces the RMS of the entire data set from $1.17\, \mps$ with the classic wavelength solution to $1.05\, \mps$.  This is equivalent to removing an independent, additive noise component of $0.52\, \mps \,(=\sqrt{1.17^2-1.05^2})$.

\begin{deluxetable}{c c c}[h]
\tabletypesize{\scriptsize}
\tablecaption{\expres\ RVs using \name\ Wavelengths \label{tab:rv}}
\tablehead{
\colhead{JD-2440000} & \colhead{RV $\mps$ } & \colhead{Error $\mps$ }
}
\startdata
 18764.4771  & 3.139  & 0.335 \\
 18764.4791  & 1.035  & 0.332 \\ 
 18764.4810  & 3.074  & 0.324 \\
 18771.4179  &-1.927  & 0.342 \\
 18771.4196 & 2.688 &	0.357 \\
          & \(\vdots\)  & \\
\enddata
\tablecomments{The full data set is available online}
\end{deluxetable}

We conducted a direct test of a classically-generated wavelength solution with \name -wavelengths on four other data sets.  All targets showed a reduction in or comparable RV RMS.  The results from these data sets can not be interpreted as directly as with \target, though, due to larger contributions from stellar variability, known planets, etc.  As completely mitigating these different effects is out of scope for this paper, we focus here on the results with \target.

\section{Choose Your Own Implementation} \label{sec:choices}
We have described and tested only one implementation of \name.  Using PCA and an interpolated wavelength solution is a statistically straight-forward step towards a complete hierarchical and non-parametric wavelength model.  It is possible to upgrade both the denoising and wavelength solution steps to true models.  It is also possible, of course, to implement either step individually.  A hierarchical framework can be used to simply denoise the lines before they are fit to a parametric model, and a non-parametric model can be used on lines that have not been denoised.

For the dimensionality reduction and denoising, the PCA could be replaced by a probabilistic PCA model or other probabilistic linear reduction methods, such as heteroscedastic matrix factorization (HMF)\citep{tsalmantza2012}.  It is also possible to move to non-linear reduction methods, like a Guassian process latent-variable model, an auto-encoder, a normalizing flow \citep[e.g.][]{kramer1991,woodbridge2020}.  Using a non-linear denoising model could enable \name\ to capture large-scale changes as well as small variations in calibration state.

The wavelength solution could also move past interpolation.  For example, a Gaussian process could be used that is constrained to ensure monotonicity.  Replacing each step with a model will allow for full, hierarchical Bayesian inference.  This means the uncertainty from wavelength calibration could be completely marginalized out.  Doing so will have the largest impact if the wavelength calibration is a significant fraction of the error budget.

The implementation of \name\ presented here, using PCA for denoising and interpolating a wavelength solution, uses various global variables and methods that we believe are or are close to optimal for constructing a high-fidelity wavelength solution.  The following subsections will describe each choice and the associated decision-making process.

\subsection{Dimensionality of the Calibration Space, $K$}
\begin{figure*}[t]
\centering
\includegraphics[width=\textwidth]{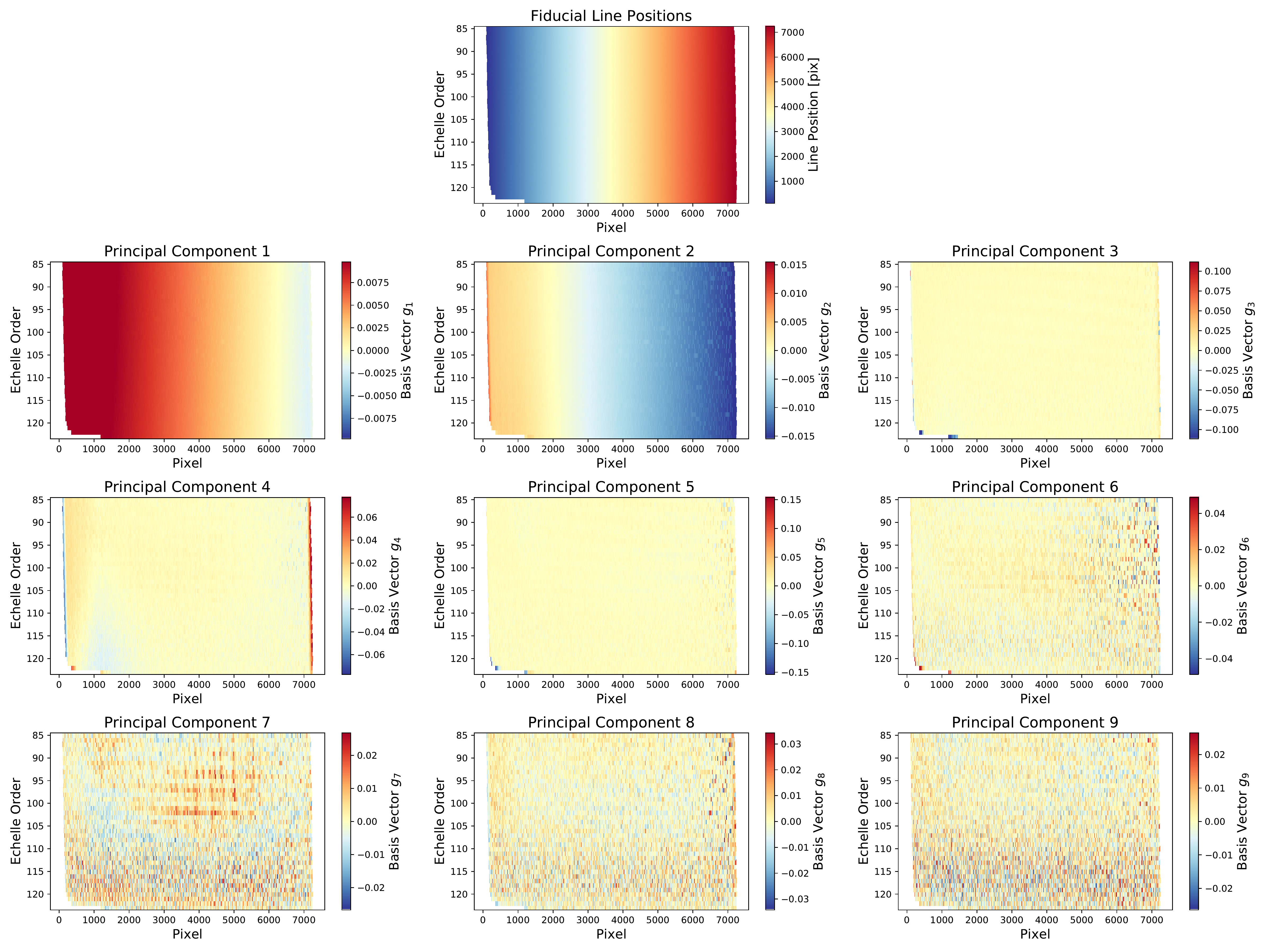}
\caption{\textbf{Top}:  The fiducial calibration of the spectrograph, i.e. the mean line positions for each line throughout the epoch of stability.  \textbf{The following $3 \times 3$ grid of plots} show the first nine principal components constructed using LFC lines.  These principal components represent the basis vectors along which the calibration of the spectrograph can deviate from the fiducial calibration.  For each principal component, or basis vector, each calibration line is plotted according to its echelle order and x-pixel and colored by the value of the basis vector for that line.  Principal components beyond the sixth one become steadily more dominated by noise.}
\label{fig:pcLfc}
\end{figure*}

\begin{figure}[b]
\centering
\includegraphics[width=.45\textwidth]{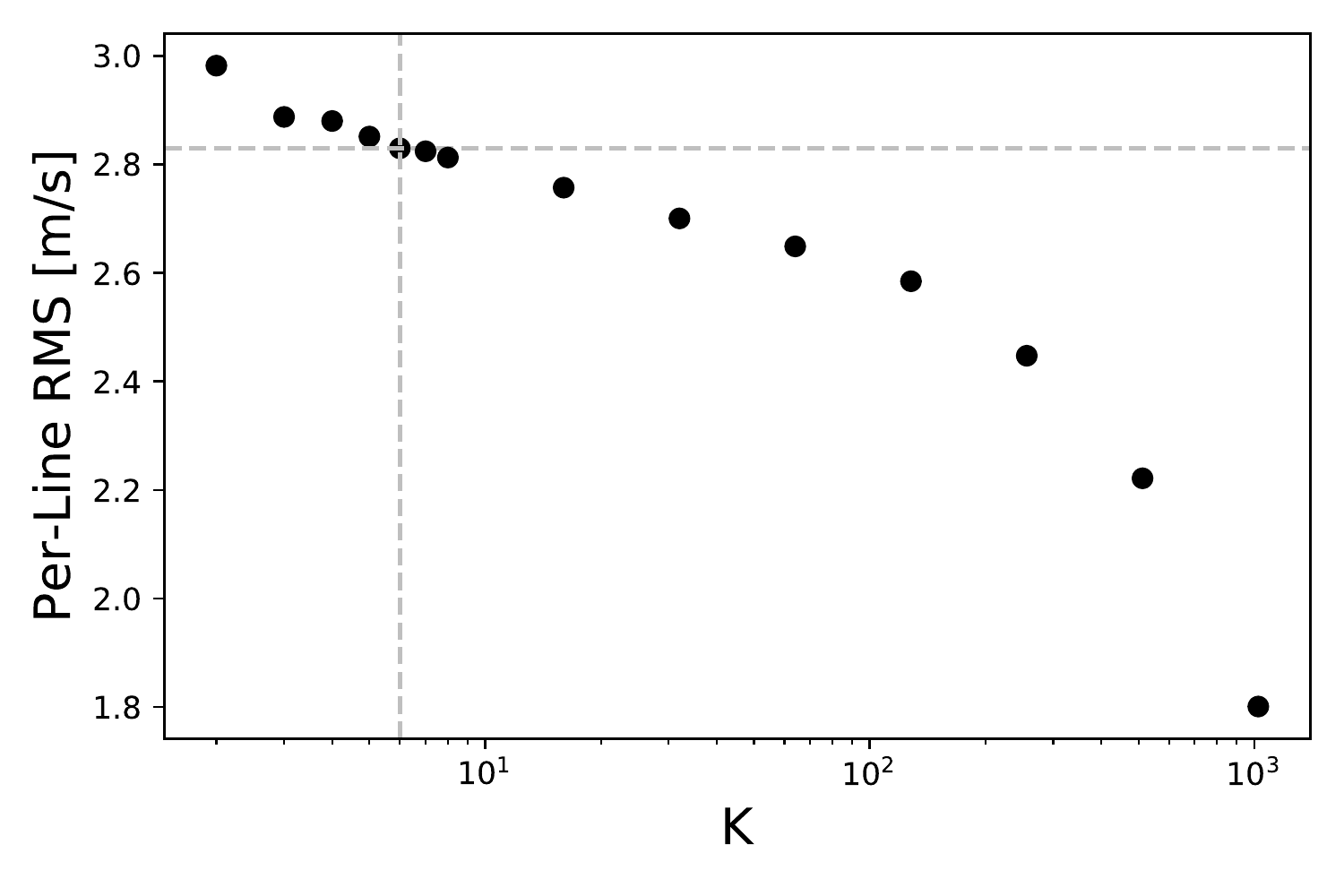}
\caption{Per-line RMS of returned wavelength models for different values of $K$.  Per-line RMS, as defined in Equation \ref{eq:rms}, provides a measure of the accuracy of a wavelength model.  There is a dotted, vertical line at $K=6$, and a dotted, horizontal line is the RMS for $K=6$.  The improvement at around $K=6$ plateaus.}
\label{fig:kvals}
\end{figure}
\label{sec:choice_k}
The dimensionality of the calibration space within which the spectrograph lives is represented by $K$.  In practice, it is the number of principal components, or basis vectors, used to reconstruct the de-noised line centers.  $K$ needs to be large enough so that all variability in the spectrograph is captured.  Too large, however, and the reconstruction will begin to incorporate noise, thereby defeating the purpose.

Figure \ref{fig:pcLfc} shows the fiducial calibration state and the first 9 principal components constructed using LFC lines, which represent deviations from the fiducial calibration state.  There is clear structure in the first and second principal components.  Components three through six show smaller or more localized structure.  Components three and four have aberrant behavior on the edges of the two bluest echelle orders, where lower signal results in more variation in the line fits.  Later principal components become dominated by noise and show less coherent structure.

In deciding a $K$ value, we ran denoising tests, as described in Section \ref{sec:tests}, for $K$ values spanning from 2 to 512.  The resultant per-line RMS for each test is plotted in Figure \ref{fig:kvals}.  One would expect the returned wavelengths to get increasingly more accurate with larger $K$ until components that represent only noise are incorporated.  Residuals might then get worse before ultimately starting to get better again with large $K$, which marks when the model starts over-fitting the data.  Though the returned RMS never gets worse, we find that the improvement plateaus between $K=6$ and $K=128$.  Comparisons of wavelengths returned by a $K=6$ model vs. a $K=32$ model show significant differences in less than 10 bluer lines, which are known to have greater error and variance in their measured line positions.  We therefore settled on a $K$ value of six.

\label{sec:choice_avt}
\begin{figure*}[t]
\centering
\includegraphics[width=\textwidth]{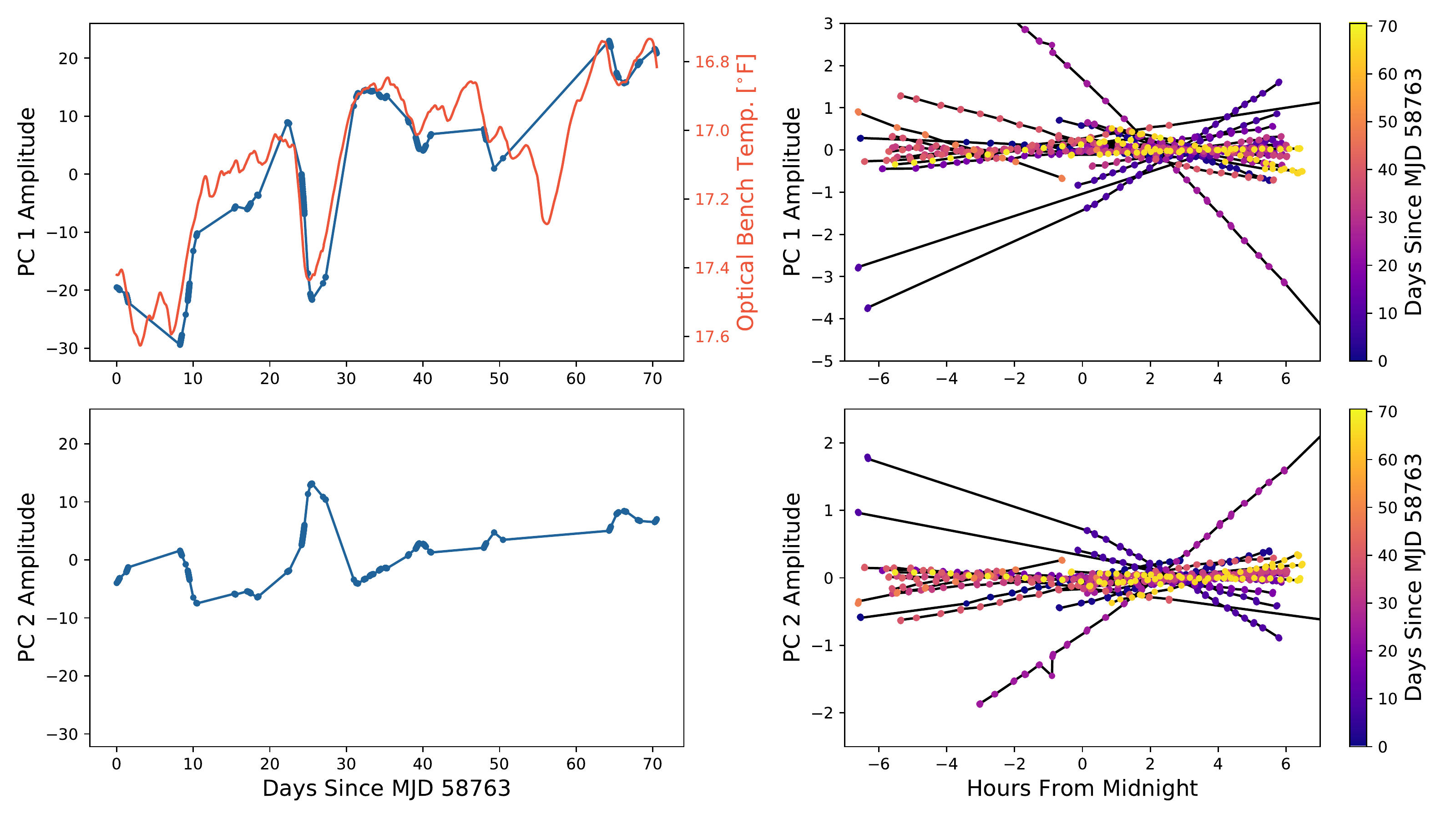}
\caption{Amplitude of the first two principal components shown as a function of time (left) or hour from midnight (right).  The top row of plots shows the amplitudes for the first principal component while the bottom row shows the amplitudes for the second principal component.  Lines show the result of a linear interpolation.  In the top, right plot, the temperature of the optical bench is also plotted in orange.  In the right plots, the principal component amplitudes for each night have been artificially offset by the median amplitude per night.  All days are therefore roughly on the same scale, but the y-axis is different from the left plots.  In the right column, points are colored by the MJD of each exposure.}
\label{fig:nightlyVariation}
\end{figure*} 

\subsection{Interpolation of Calibration State to Science Exposures}
Figure \ref{fig:nightlyVariation} shows the amplitude of the first and second principal component with respect to time on the left.  Though there exists a complex overall shape to the amplitudes with respect to time, a clear linear trend exists within each night.  This is shown by the right plots in Figure  \ref{fig:nightlyVariation}.  As the beginning-of-night and end-of-night calibration sets always include LFC exposures, we use a simple linear interpolation to interpolate principal component amplitudes with respect to time.

The choice in interpolation method can help guide how many wavelength calibration images are truly needed.  It is unnecessary to take calibration images at times where the same information can be reconstructed at the desired precision by a well-chosen interpolation scheme.  For example, with the \expres\ data shown here, it is clear that nightly calibration images are needed, but for a linear trend, only two calibration images throughout the night are strictly required.

We also tested an implementation of \name\ where the $K$ principal components within a night were fit to a cubic with respect to time rather than linearly interpolated.  This emulates the current, polynomial-based wavelength solution implemented in the \expres\ pipeline, where polynomial fits to calibration files are interpolated to science exposure by fitting polynomial coefficients with respect to time to a cubic.  We found that using a cubic in place of linear interpolation returns comparable RV RMS for most targets, though appears to do better when a night has sparse calibration data.  This suggests that the nightly behavior of \expres\ with respect to time is well described by a cubic function, but LFC exposures are typically taken with enough frequency that a linear interpolation provides a good approximation (see Figure \ref{fig:nightlyVariation})

The amplitudes $a_{n,k}$ can also be interpolated with respect to any good housekeeping data, not just time.  Best results will come from interpolating with respect to whatever is most strongly correlated with the calibration state of the spectrograph.  For example, with \expres, which is passively temperature controlled, the returned amplitudes $a_{n,k}$ were extremely correlated with the optical bench temperature, as shown in the top-left plot of Figure \ref{fig:nightlyVariation}, suggesting it would also be possible to interpolate the amplitudes with respect to temperature.

Another pertinent example would be a spectrograph that is mounted on the telescope, and therefore moves with the telescope.  In this case, it may be important to interpolate at least in part with respect to the position of the telescope, which enables the resultant calibration to incorporate the gravitational loading experienced by the spectrograph.

\begin{figure*}[t]
\centering
\includegraphics[width=\textwidth]{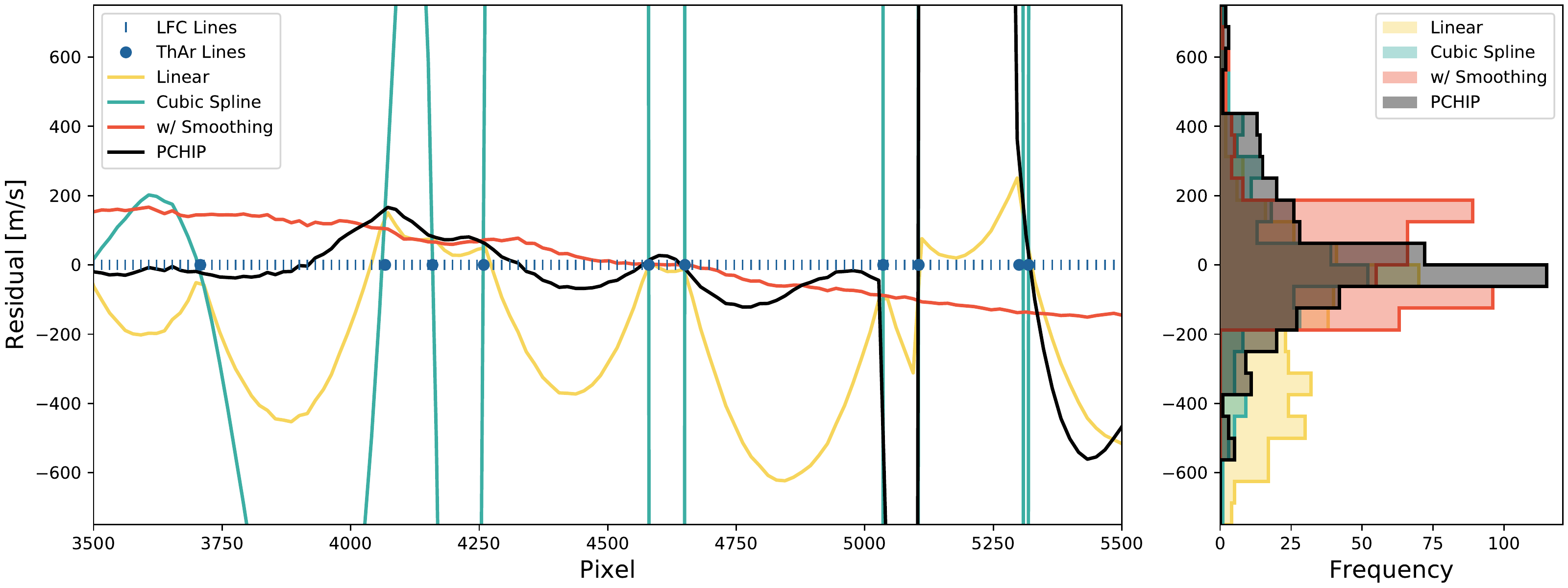}
\caption{Residuals from different interpolation schemes over pixels in echelle order 100.  ThAr lines, shown as blue circles, are used to construct a wavelength solution that is then evaluated at each LFC line, shown as blue vertical lines.  The residuals of each wavelength solution for a subset of the order is shown on the left.  Histograms of the residuals for each method for the complete order is shown on the right.  Note: there is a blended ThAr line at approximately pixel 5300, the right most ThAr lines plotted.}
\label{fig:xinterp}
\end{figure*}

\subsection{Interpolation of Wavelengths with Respect to Pixel}
\label{sec:choice_wvp}
In the implementation described and tested by this paper, interpolation of wavelengths over pixel  is done order-by-order using a Piecewise Cubic Hermite Interpolating Polynomial (PCHIP) interpolator.  This interpolation incorporates the flexibility needed to model the changing dispersion of the spectrograph across an echelle order along with any detector defects while also enforcing monotonicity, which we know must be true across any one echelle order.

A simple linear interpolation would give erroneously low values everywhere.  Due to the dispersion intrinsic to echelle spectrographs, the wavelength change between pixels grows greater with greater wavelengths, even within an order.  This means that the function of wavelength vs. pixel across an echelle order will always be monotonically increasing and concave down everywhere.

Though less of an issue with LFC lines, a more classic cubic spline interpolation can run into issues with arc lines, which are irregularly spaced or even blended.  Close lines appearing at virtually the same pixel location but with different wavelengths could coerce a cubic spline into a very high slope.  This is demonstrated by the green line in Figure \ref{fig:xinterp}, which shows the results of interpolating between ThAr lines rather than LFC lines.  A line blend appears at approximately pixel 5300, causing the spline to twist to nearly vertical to account for both points.  This leads to huge deviations from the correct wavelengths around this line blend as the extraneously high slope of the spline is accounted for.

These huge digressions can be avoided by allowing for some smoothing in the interpolation.  In Figure \ref{fig:xinterp}, we show an example in orange using SciPy's implementation of a univarate spline.  While the result appears to follow the calibration lines much better, the smoothing ultimately causes larger residuals that are spatially correlated (Fig. \ref{fig:xinterp}, right).  In all echelle orders, the edges are overestimated while the middle will be underestimated, shown by the flattened shape of the histogram of residuals.  The resultant wavelength solution is underestimating the curvature of the pixel-wavelength relation, giving rise to similar issues as with an inflexible, parametric wavelength solution.  Introducing this smoothing parameter is enforcing a smoothness we have no reason to believe is true of the data, thereby re-introducing one of the problems with parametric models.

We instead turn to the PCHIP algorithm, which damps down huge deviations in the traditional cubic spline by requiring the resulting interpolated function be monotonic.  Monotonicity is a constraint we know must be true for any one echelle order.  Though the PCHIP interpolator shows a similar issue as a classic cubic spline around the ThAr line blend at pixel 5300, the effect is much smaller and affects fewer pixels.  Figure \ref{fig:xinterp}, right, shows that using the PCHIP interpolator returns the lowest spread residuals.  

There likely exists an even more fitting model between an overly-constrained polynomial fit or a completely free spline interpolation.  For example, there has been success interpolating wavelengths with respect to pixel using a segmented polynomial in the dispersion direction, especially when tuned to known detector defects \citep{milakovic2020}.  Stiffer, more constrained splines or carefully chosen knot position may afford the perfect marriage of freedom and constraint that will better describe wavelength changes with pixel.

\section{Application to Other Spectrographs}
\label{sec:others}
We focus on an \eprv\ use-case here because there is a strong need for wavelength calibration in the \eprv\ community.  \expres\ is representative of the newest generation of \eprv\ spectrographs, and an LFC provides stable, dense calibration lines with known wavelengths, ideal for \name.  The applicability of \name\ to any one instrument is a detailed question of the kind of variance experienced by the spectrograph and calibration sources available, but we hope to provide some approximate benchmarks and diagnostics here.

Implementing \name\ will require an existing library of calibration exposures that span the range of calibration space accessible by the spectrograph, or at least the space accessed by the science exposures being calibrated.  A new set of calibration exposures would be needed for any changes to the instrument that could significantly change the accessible calibration space.  While there is no exact cutoff for how many exposures are needed, the number is certainly much larger than $K$, the dimensionality of the calibration space.  More calibration exposures will help with denoising .  The number of calibration exposures required throughout a night will depend on how much the instrument varies throughout the night as well as the chosen interpolation scheme, as mentioned in Section \ref{sec:choice_avt}.

As an empirical assessment of the needed line density, we removed LFC lines from the \expres\ data and calculated the per-line RMS of the returned wavelengths for the removed lines.  These tests are similar to the interpolation test explained in Section \ref{sec:tests}, in that a fraction of lines are systematically removed from the analysis.  An increasing fraction of lines are removed to simulate different line densities.  For these line density tests, though, we are also implementing denoising unlike the pure interpolation test of Section \ref{sec:tests}.

\begin{figure}[b]
\centering
\includegraphics[width=0.45\textwidth]{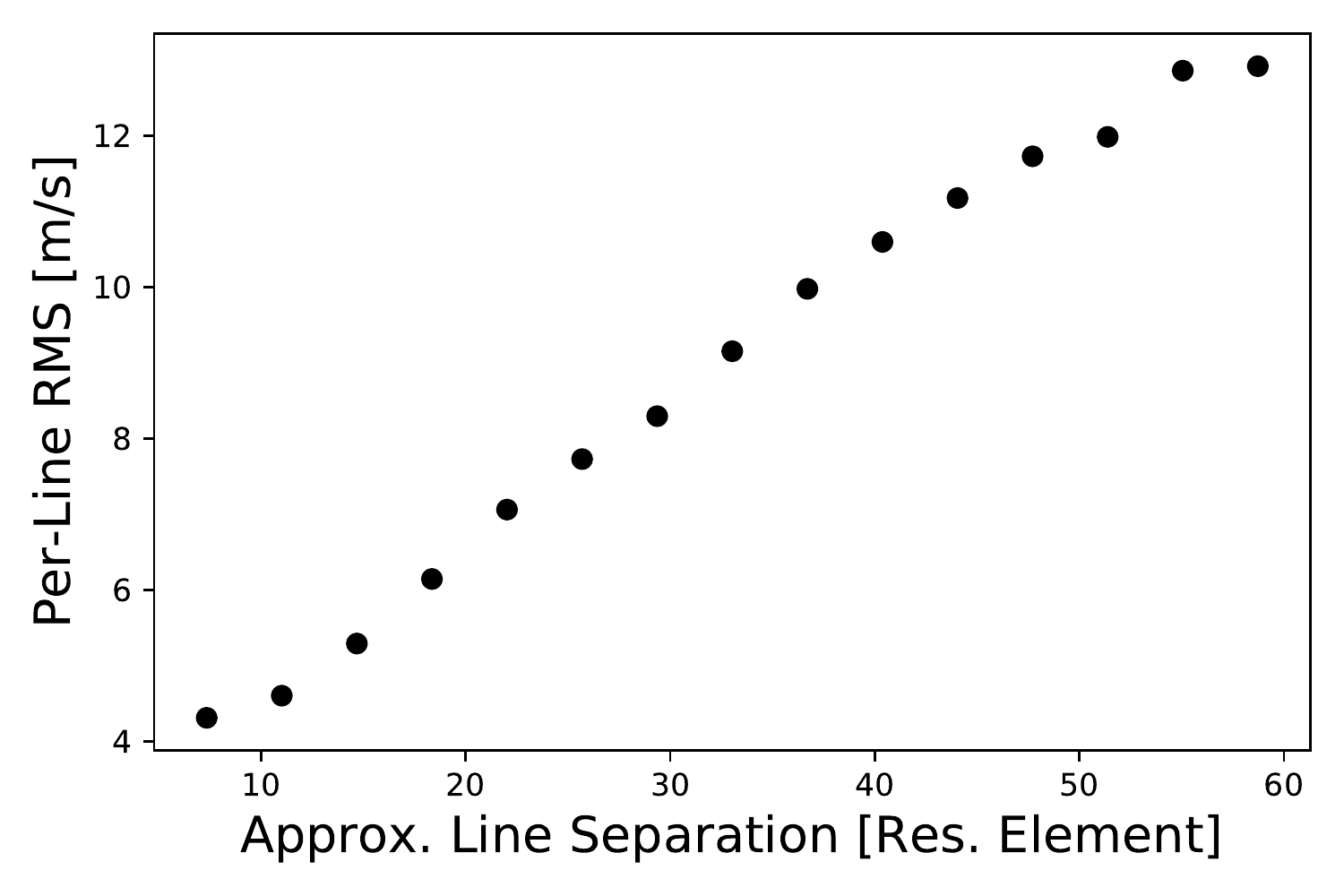}
\caption{Per-line RMS as a function of spacing between lines in units of resolution element.  The average line spacing is calculated using the average distance between LFC lines across its wavelength range.}
\label{fig:density}
\end{figure}

Figure \ref{fig:density} plots the scatter of residuals of the wavelengths returned by \name\ as a function of separation between lines in units of resolution element.  It gives an approximate estimation of the expected error between lines of different separation.  Note, however, that the stability of the lines and their measured line centers quickly becomes the dominant error source in calibration lamps over the separation between lines.  Additionally, in this assessment, the lines remain uniformly spaced.  The needed line density depends on the resolution of a given spectrograph, the needed precision of the returned wavelength solution, and the chosen interpolation scheme.

This implementation of \name\ to \expres\ data interpolates calibration information to the science data using surrounding calibration exposures.  Simultaneous calibration data can be used to reinforce the determined calibration state of an instrument at the time of science exposures.  This simultaneous calibration data can come from calibration light shined through a reference fiber or any meta-data (e.g. temperature, pressure, time, etc.) that correlates with the calibration.  For example, we have seen that the calibration state of \expres\ is correlated with the temperature of the optical bench, even with the optical bench temperature varying by less than one degree.  This correlation is not seen in the RVs, suggesting the changes with temperature are calibrated out at the wavelength stage.

With a simultaneous reference fiber, the position of calibration lines taken through the reference fiber can simply be concatenated to the array of line positions taken through the science fiber.  Both sets of lines will then be used when the low-dimensional basis is constructed.  This allows the simultaneous calibration information to contribute to constructing the complete calibration space of the spectrograph and pinpoint where the spectrograph is in that calibration space for any exposure.

The calibration for any science exposure with a simultaneous reference can be determined by finding the amplitude of each basis vector that most closely recreates the calibration line positions through the reference fiber.  These amplitudes can then be used to recreate the calibration line positions through the science fiber as well.  This replaces the need to interpolate the basis vector amplitudes from calibration exposures to science exposures, something that is done with respect to time in the example implementation described in this paper.  The result is likely to be even more precise, as this method incorporates more data.  This method, as with all analysis involving a simultaneous reference fiber, will work only as well as the reference fiber's ability to trace changes in the main science fiber.

It is also possible to apply \name\ to etalon data with some modifications.  The simplest implementation is if the free spectral range (FSR) and therefore wavelength of each line of the etalon is well known.  The etalon data can be interpolated onto a set of fiducial lines with set wavelengths.  These fiducial lines would therefore be identifiable by echelle order and wavelength alone, with only their line positions varying with different exposures.  This returns us to the same framework as developed for the case of an LFC.  This marks the simplest implementation of \name\ on etalon data, as the uncertainty of a line's wavelength is upstream of the model rather than built in.

Incorporating the FSR as part of the \name\ model will require introducing a free parameter to capture changes in the FSR independent of variation in an instrument's calibration state.  The calibration state can then be described with respect to mode number, which will be used to uniquely identify a calibration line across exposures rather than wavelength.  The FSR is then used to determine how the mode number of each line maps to wavelength for a given exposure.  The FSR must not vary so much that the change in this mode-number-to-wavelength mapping becomes non-linear.  This model would require a simultaneous reference or other housekeeping data that can be used to determine the FSR for every exposure.

In terms of dimensionality reduction, most physical systems should have only a few dominant axis along which they vary, meaning \name\ should be adaptable to a wide range of instrument designs.  With PCA, this can be tested by plotting the amplitude of the returned principal components, which should fall quickly after a few components.  It should be noted that this only provides a measure of the variance in the PCA space, and is not an explicit test of the end-to-end variation in the resulting model.  This condition is therefore necessary but not sufficient if implementing \name\ with PCA.

It could still be possible to run \name\ on a spectrograph that has a high-dimensional calibration space, meaning a large number of basis vectors are required to capture all the variance in the spectrograph.  In this regime, there is always the risk of over-fitting.  Regularizing the principal component amplitudes, for example insisting the amplitudes for higher principal components be smaller, can help to return reasonable models \citep{formanmackey2015} .  Within such a framework, \name\ may still deliver good results.

For the results presented here, the data was broken up into different eras of stability based on where the principal component amplitudes showed huge deviations.  This was done visually, though there exists many change-point detection algorithms that could be used \citep{aminikhanghahi2017}.  There is a trade-off in including more exposures between introducing greater variation, but also more data to offer constraints that may be optimized.  Here, an era of stability was chosen  manually in order to focus on separating out time-domain intervals in which the data is relatively homogeneous, e.g. most exposures show the same calibration lines.  Homogeneity is, of course, implicitly required when implementing PCA.  Different denoising models will be able to account for different amounts of stability or lack thereof.

Lastly, we caution that \name\ is extremely sensitive to upstream changes that may effect the line centers.  For example, PCA is good for detecting variability, but is agnostic to the source of the variability.  This is why the principal components shown in Figure \ref{fig:pcLfc} exhibit errant values for bluer LFC lines, which are lower signal and therefore exhibit more variation in their fitted line centers.  It is essential that the line positions being fed to \name\ capture only the changes in the spectrograph's calibration state, not potential errors in the fitted line centers.

\section{Discussion} \label{sec:discussion}
We show that \name\ returns a lower per-line RMS than classic, parametric methods by a factor of 5 (Section \ref{sec:tests}).  The residuals were also smoother, exhibiting less spatial correlation (Figure \ref{fig:resid2d}).  Using \name\ wavelengths reduced the RMS in RVs of HD 34411 from $1.17\, \mps$ to $1.05\, \mps$ (Section \ref{sec:test-rv}).

In implementing \name\ on \expres\ data, we have successfully constructed a model of \expres's accessible calibration space, confirming that \expres\ truly is an instrument with low-degrees of freedom.  \Name\ does not make any claims about what variability each basis vector represents.  Those interested in interpreting the variability are encouraged to investigate how the amplitude of the different vectors varies with different housekeeping data to find their source.

Starting with a list of calibration lines with known wavelengths and well-fit line centers for each calibration exposure, \name\ will de-noise and interpolate the given lines into a full wavelength solution.  \Name\ leverages the stability of contemporary EPRV instruments and high density of lines made available by new calibration sources, such as LFCs and etalons, to achieve more accurate wavelengths.  \Name\ therefore assumes dense enough calibration lines to properly constrain a non-parametric wavelength model, and that the instrument has low degrees of freedom.

Denser calibration lines allow us to move to more flexible wavelength models, which can then account for non-smooth features in the wavelength solution.  Stabilized spectrograph hardware makes it more likely that the calibration space of the instrument is low-dimensional.  All calibration images in a given generation of stability can therefore be used to constrain the accessible calibration space of the spectrograph as well as where in that calibration space the spectrograph lies.  We have described only one, fairly simplistic implementation of \name\ here.  There are many options for both the de-noising and interpolation steps, as mentioned in Section \ref{sec:choices}.

An advantage of this implementation of \name, where PCA is applied to line positions from all LFC exposures, is the ability to isolate exposures that exhibit errant variation, which is typically associated with flawed exposures.  This allowed us to quickly vet for problematic LFC exposures, which otherwise would have required visual inspection of all 1200+ LFC exposures.  In a classic framework where each calibration exposure is treated independently, these aberrant exposures would likely have persisted undetected and are liable to sway the resultant wavelength solutions for at least an entire night.

On the other hand, PCA captures all variance, regardless of source.  Though \name\ endeavors to capture only variation in the instrument, the PCA is also sensitive to uncertainties and failures in the upstream line-position fitting.  For example, we have seen that lower-signal lines that are harder to fit faithfully will have greater variety in returned line positions, which is in turn captured by the PCA.  In this sense, \name\ is actually a model of not just the instrument, but all the upstream choices used to drive live positions.  High-fidelity line positions are essential to ensure the PCA is capturing variations in just the spectrograph's calibration rather than changes in how well a line can be fit or other effects.

Along those lines, we caution that with any wavelength solution, there is a perfect degeneracy between what is defined as the ``position of the line'' and the resultant wavelength solution.  If, for example, a cross correlation method is used to extract RVs from the data, a systematic difference may be introduced depending on what exactly is defined to be the line position, whether it be the mode of a Gaussian fit, the first moment of a complicated PSF, or the output of some particular peak-finding algorithm, etc.  In principle, the best way to mitigate this uncertainty would be to use a calibration source that looks like a star.

Compared to traditional methods, which involve fitting high-order polynomials, \name\ has several useful statistical properties.  \Name\ is technically robust to localized issues that arise from either the calibration source or the pipeline used to return line positions.  With an interpolated wavelength model, one errant line position will only effect the resultant wavelength model out to the close, neighboring lines.  The effect of an outlier is diminished and kept localized.  In contrast, an entire parametric fit will be affected by a single errant line, changing the wavelength solution for the entire exposure for a 2D fit.  Through de-noising and outlier rejection, \name\ adds additional robustness against erroneous line positions.

The locality of the interpolation caries other benefits as well.  Manufacturing artifacts in the detector or other optical elements can lead to non-smooth structure in the wavelength solution that can not be captured by polynomials or other smooth functions (see \figurename~\ref{fig:resid2d}).  An interpolated model introduces greater flexibility, enabling the model to account for such high-order effects.  As discussed in Section \ref{sec:choice_wvp}, there are better and worse interpolators for the task, which may differ for different instruments and different calibration sources.  Instead of using an interpolator at all, there might be better results from implementing something more sophisticated, such as a kernel method or a Gaussian process with a kernel adapted for the specifics of an instrument.  There is in principle an enormous number of non-parametric methods to explore, which we leave outside the scope of this paper.

Similarly, PCA is just one of many possible dimensionality-reduction methods.  We chose to implement \name\ using PCA here for simplicity and computational tractability.  PCA is a good option because the instrument changes here are small enough that a linear model is an approrpiate representation of the changes.  If \name\ was updated to a full probabilistic model, the PCA along with the interpolation model would have to be upgraded to something with better probabilistic properties.  Other, nonlinear denoising methods may be more robust to large changes, allowing all calibration images ever taken with an instrument to be used to construct the accessible calibration state regardless of hardware adjustments.  Further discussion of other implementations of \name\ can be found in Section \ref{sec:choices}.

\Name\ can be applied to any data that contains information about the calibration state of the spectrograph (see Section \ref{sec:others}).  For example, though LFC and ThAr exposures are used as an example in this paper, \name\ would work similarly for an etalon or any other arc lamp with a list of lines and assigned wavelengths.  Simultaneous calibration information can easily be accounted for by simply including the line position information from the simultaneous reference when constructing a low-dimensional basis of the instrument's calibration space.

Once we have defined a calibration space that captures all possible degrees of freedom for a stabilized spectrograph, there are many options for pinpointing where the spectrograph is located within that calibration space.  Good housekeeping data, such as temperature or pressure could be used in addition to or instead of time (as mentioned in Section \ref{sec:choice_avt}).  Telemetry that is seen to be correlated with the calibration state of the spectrograph can even be added to the data used to construct the low-dimensional basis.  Furthermore, all exposures taken with the spectrograph in principle contains information about the calibration state of the spectrograph.  Theoretically, tellurics, lines in science exposures, or just the trace positions themselves could also be used to determine an instrument's calibration state, thereby providing free simultaneous calibration information.

\Name\ is designed and optimized for extreme-precision, radial-velocity spectrographs.  In the battle to construct a high-fidelity data pipeline for extreme-precision radial-velocity measurements, we have shown that \name\ represents a step towards mitigating the error from wavelength calibration, as demonstrated by tests using \expres\ data (Section \ref{sec:tests}).  Though the focus was on \eprv\ instruments here, \name\ should be largely applicable to nearly any astronomical spectrograph.

\software{SciPy library \citep{scipy}, NumPy \citep{numpy, numpy2}, Astropy \citep{astropy:2013,astropy:2018}.}

\facilities{LDT}

\acknowledgements
We  gratefully  acknowledge  the  anonymous  referee whose  insightful  and  well thought out  comments  significantly improved the paper.  We thank the EXPRES team for their work on the instrument, software, and pipeline.  We thank Dan Foreman-Mackey for illuminating discussion.  The data presented here made use of the Lowell Discovery Telescope at Lowell Observatory. Lowell is a private, non-profit institution dedicated to astrophysical research and public appreciation of astronomy and operates the LDT in partnership with Boston University, the University of Maryland, the University of Toledo, Northern Arizona University and Yale University. We gratefully acknowledge ongoing support for telescope time from Yale University, the Heising-Simons Foundation, and an anonymous donor in the Yale community. We especially thank the NSF for funding that allowed for precise wavelength calibration and software pipelines through NSF ATI-1509436 and NSF AST-1616086 and for the construction of EXPRES through MRI-1429365.  LLZ gratefully acknowledges support from the National Science Foundation Graduate Research Fellowship under Grant No. DGE1122492.

\bibliography{paper}

\end{document}